\documentclass[pdflatex,sn-mathphys-num]{sn-jnl}

\usepackage{graphicx}%
\usepackage{multirow}%
\usepackage{amsmath,amssymb,amsfonts}%
\usepackage{amsthm}%
\usepackage{mathrsfs}%
\usepackage[title]{appendix}%
\usepackage{xcolor}%
\usepackage{textcomp}%
\usepackage{manyfoot}%
\usepackage{booktabs}%
\usepackage{algorithm}%
\usepackage{algorithmicx}%
\usepackage{algpseudocode}%
\usepackage{listings}%
\usepackage{booktabs}  
\usepackage{adjustbox}
\usepackage{placeins} 
\usepackage{caption}
\usepackage{lineno}
\usepackage[title]{appendix}

\theoremstyle{thmstyleone}%
\theoremstyle{thmstyletwo}%

\theoremstyle{thmstylethree}%

\begin{document}
	
\title[Article Title]{The Radioactive Background of the JUNO Calibration System}

\author[1]{\fnm{Rui} \sur{Li}}
\author[1]{\fnm{Youhui} \sur{Yun}}
\author[1]{\fnm{Zhangming} \sur{Chen}}
\author[1]{\fnm{Junting} \sur{Huang}}
\author[2]{\fnm{Jiaqi} \sur{Hui}}
\author[1]{\fnm{Haojing} \sur{Lai}}
\author[1,2,3]{\fnm{Jianglai} \sur{Liu}}
\author[4]{\fnm{Yankai} \sur{Liu}}
\author*[1]{\fnm{Yue} \sur{Meng}}\email{mengyue@sjtu.edu.cn}
\author[2]{\fnm{Akira} \sur{Takenaka}}
\author[1]{\fnm{Ziqian} \sur{Xiang}}
\author[1]{\fnm{Feiyang} \sur{Zhang}}
\author[1]{\fnm{Ping} \sur{Zhang}}
\author[4]{\fnm{Qingmin} \sur{Zhang}}
\author[2]{\fnm{Tao} \sur{Zhang}}
\author[2]{\fnm{Yuanyuan} \sur{Zhang}}

\affil[1]{\orgdiv{School of Physics and Astronomy, MOE Key Laboratory for Particle Astrophysics and 
Cosmology, Shanghai Key Laboratory for Particle Physics and Cosmology}, \orgname{Shanghai Jiao Tong 
University}, 
	\orgaddress{\city{Shanghai}, \postcode{200240}, \country{China}}}

\affil[2]{\orgname{Tsung-Dao Lee Institute, Shanghai Jiao Tong University}, 
	\orgaddress{\city{Shanghai}, \postcode{201210}, \country{China}}}

\affil[3]{\orgdiv{New Cornerstone Science Laboratory}, \orgname{Tsung-Dao Lee Institute, Shanghai 
Jiao Tong University}, 
	\orgaddress{\city{Shanghai}, \postcode{201210}, \country{China}}}
\affil[4]{\orgdiv{School of Nuclear Science and Technology}, \orgname{Xi’an Jiao Tong University}, 
	\orgaddress{\city{Xi’an}, \postcode{710049}, \country{China}}}
	
\abstract{
	The Jiangmen Underground Neutrino Observatory (JUNO) experiment is a reactor antineutrino 
	detector employing 20~kton of ultra-pure liquid scintillator to determine the neutrino mass 
	ordering and to precisely measure oscillation parameters. 
	The total singles background rate from radioactivity is required to be below 10~Hz in the 
	energy range of 0.7--12~MeV within the fiducial volume for reactor neutrino analysis. 
	The calibration system is designed to characterize the detector energy and position responses, 
	while several of its components are located close to the target and may contribute to the 
	background budget. 
	Therefore, extensive material screening and selection are required to construct a 
	low-background calibration system and to ensure that its contribution remains within the design 
	requirements.
	In this work, a comprehensive study of the radioactive background induced by the calibration 
	system is presented, including material radioactivity measurements using high-purity germanium 
	detectors and neutron activation analysis techniques, detailed Monte Carlo 
	simulations to evaluate the background,and comparisons with in-situ detector data to validate 
	the predictions.
	In this data analysis, dedicated spatial selection methods are developed to isolate 
	calibration-related contributions and to suppress the liquid scintillator background. 
	The total radioactivity contribution from the calibration system is estimated to be less than 
	76~mHz, which satisfies the requirement of 200~mHz (2\% of the total background budget). 
	The results from in-situ data are found to be consistent with the expectations based on 
	material assay and simulation within uncertainties. 
	These results demonstrate that the calibration-induced background is well understood, in 
	agreement between data and simulation, and negligible for reactor antineutrino measurements in 
	JUNO.
}

	
	\maketitle
	
	\section{Introduction}
	The Jiangmen Underground Neutrino Observatory (JUNO) is an experiment located at an underground 
	laboratory with 650~m mountain shielding in Jiangmen City, Guangdong Province, China. With 
	$\sim$53~km baselines from both Yangjiang and Taishan nuclear power plants 
	(NPPs)~\cite{yellow-book},  it aims to determine the neutrino mass ordering and precisely 
	measure neutrino oscillation parameters.
	In its central detector (CD), 20~kton ultra-pure liquid scintillator (LS) is filled into an 
	acrylic spherical vessel with an inner diameter of 35.4 m and a thickness of 12 cm. There are 
	about 17,600 20-inch and 25,600 3-inch photomultipliers (PMTs) facing to the LS to collect 
	photons. The CD is located in a cylindrical ultra-pure water pool to effectively shield 
	external radioactivities. The water pool (a Cherenkov detector) and top tracker(TT) are also 
	used to tag cosmogenic background.
	
	The reactor antineutrino could be detected and identified in JUNO via so-called inverse beta 
	decay (IBD) reaction, $\bar{\nu}_{e}+p \rightarrow e^{+}+n $.
	As a prompt signal, the positron deposits energy in the LS in a short time ($\sim$ns level) and 
	then annihilates with an electron to produce two $0.511$ MeV gamma rays,
	and the neutron could be mainly captured on proton in the LS within about 200 $\mu$s, 
	producing a delayed gamma signal. IBDs can be identified by such prompt-delayed coincidences. 
	However, MeV scale radioactive background can mimic either the prompt or delayed signals to 
	produce accidental background. 
	In Ref.~\cite{PPNP}, the requirement of single rate of background is less than 10~Hz in 
	the fiducial volume with radius less than 17.2 m, and the calibration system is required to 
	contribute no more than $2\%$ of the total singles background~\cite{Sisti2020LowBackground}, 
	posing significant challenge to background control. \\

	\section{The JUNO calibration system}
\begin{figure}[!h]
		\centering
		\includegraphics[width=3in]{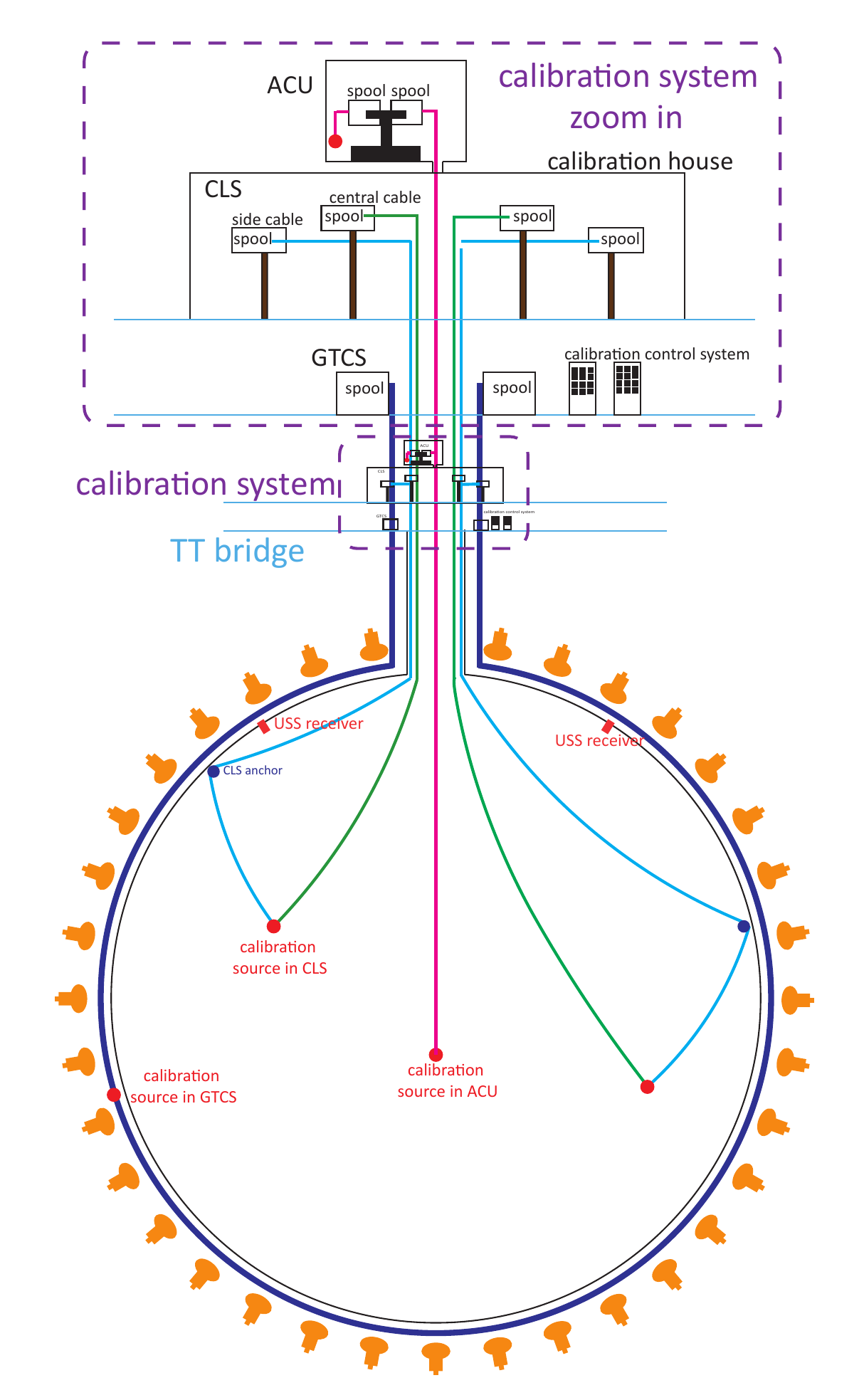}
		\caption{The overview of the calibration system, including Automatic Calibration Unit, 
		Cable Loop Systems, and the Guide Tube Calibration System.}
		\label{fig:calibsystem}
	\end{figure}

    The JUNO calibration system is designed to correct energy non-linearity and detector 
    non-uniformity, with multiple calibration sources deployed to designated positions within the 
    detector~\cite{calib_strategy_paper}.
	The calibration system includes four sub-systems and a positioning sub-system, as shown in 
	Fig.~\ref{fig:calibsystem}, and is coupled to the CD through a chimney structure that connects 
	the calibration house to the top of the detector. In what follows, we discuss the basic 
	geometry of each sub-system, with detailed material information summarized in 
	Table~\ref{tab:Radio_measure}.
	
	\indent
	The Automatic Calibration Unit (ACU) deploys the given sources along the central axis of the 
	CD~\cite{JUNO-ACU-paper}. The rate of the three radioactive sources inside is $\sim$100 Bq 
	level each~\cite{Zhang:2018yso,Takenaka:2024ctk}. During normal physics data taking, the entire 
	ACU is located above the CD with 8.85~m height chimney and 2 m height calibration house, and 
	all cables will be retrieved to the ACU chamber, the background contribution from the ACU is 
	negligible.\\
	\indent
    The Cable Loop System (CLS) deploys calibration sources to off-axis positions in a vertical 
    plane~\cite{zhangyy-paper}. There are two asymmetric CLSs as illustrated in 
    Fig.~\ref{fig:calibsystem}. The calibration source is attached at the junction of a central 
    cable and a side cable (blue wire in Fig.~\ref{fig:calibsystem}). A piece of 
    polytetrafluoroethylene (PTFE) anchor is permanently fastened on the inner surface of the 
    acrylic sphere to guide and support the side cable for each CLS. The central and side cables 
    pass through the anchor, with one end connected to the calibration source, while the other end 
    extends upward through a chimney structure of about 10~m in length, composed of acrylic and 
    stainless steel, reaching the calibration house~\cite{Hui_2025}. During calibration, both the 
    central and side cables are immersed in the LS. After calibration, the central cable is 
    retracted, while the side cable remains inside the CD, with a residual length of less than 
    71~m. The CLS cable consists of a stainless steel (SS) central core with fluorinated ethylene  
    propylene (FEP) jacket, a total diameter of 1 $\rm mm$ and an average density of 3.0 $\rm g/m$.

	
	The Guide Tube Calibration System (GTCS) consists of a PTFE tube, proximity sensors and SS 
	cables. The PTFE tube is located outside of the CD surrounding the acrylic 
	sphere in the longitude orientation. A radioactive source could be driven by cables attached to 
	both ends~\cite{GTCS}. Ten proximity sensors are used to determine the locations of the source.
	Radioactivity of the PTFE tube, sensors and cables can penetrate into the acrylic sphere and 
	deposit energy in the LS target~\cite{Guo:2021ugw}. 

	The ultrasonic receivers of the UltraSonic positioning System (USS) are fastened at the inner 
	surface of the CD to receive the ultrasonic wave from the transmitter attached to the 
	calibration source of the CLS~\cite{USS_paper}. There are 10 ultrasonic receivers in total. 
	Four of them are located along the equator of the acrylic sphere, while the remaining six are 
	positioned at a northern latitude of $58^{\circ}$, uniformly distributed in azimuth with an 
	angular spacing of $60^{\circ}$. All the receivers, cables and fasten structure are located in 
	the LS permanently.

	\section{Radioactivity measurement and material selection}
	
	To control and minimize the intrinsic radioactive background introduced by the calibration 
	system, a material screening and selection program was carried out. 
	Most materials and components were assayed using high-purity germanium (HPGe) detectors located 
	at the China JinPing underground Laboratory (CJPL),	while PTFE samples were additionally 
	measured using Neutron Activation Analysis (NAA) at INFN Milano Bicocca.

	The two techniques are complementary. HPGe measures the $\gamma$-rays emitted by daughter 
	nuclides, and the chain activities are derived under secular equilibrium within each sub-chain, 
	with $^{238}$U, $^{232}$Th, $^{40}$K, $^{60}$Co and $^{137}$Cs as the target nuclides. 
	The characteristic $\gamma$-rays used in the HPGe analysis include $^{60}$Co 
	(1173, 1332~keV), $^{137}$Cs (662~keV), $^{40}$K (1461~keV), the $^{232}$Th chain ($^{228}$Ac 
	at 338, 911, 969~keV and $^{208}$Tl/$^{212}$Pb at 239, 583, 2615~keV), the $^{238}$U early 
	sub-chain ($^{226}$Ra at 186~keV) and late sub-chain ($^{214}$Pb at 295, 352~keV and $^{214}$Bi 
	at 609, 1120, 1764~keV). NAA 
	activates and measures the parent U and Th nuclei directly, giving nuclide mass fractions (g/g) 
	that are insensitive to radon escape. Since radon diffusion may break secular equilibrium, the 
	$^{238}$U and $^{232}$Th chains are split into early (parent-to-radon) and late 
	(radon-to-stable) sub-chains and treated independently in the simulation input and the in-situ 
	analysis; the screening results in Table~\ref{tab:Radio_measure} 
	report the measured chain (HPGe) or parent (NAA) activities, which are used as simulation 
	inputs under the sub-chain equilibrium assumption.

	Since several calibration components are permanently installed inside or close to the central 
	detector, such as the CLS cable, USS receivers, and GTCS proximity sensors, 
	their intrinsic radioactivity can directly contribute to the detector background. 
	Therefore, special attention was paid to the selection of materials for these components.
	
	For key elements such as the CLS cable, multiple batches of stainless-steel wires were procured 
	from different batches and measured individually using HPGe detectors. 
	The activities of relevant radioactive isotopes, including $^{238}$U, $^{232}$Th, $^{40}$K, 
	$^{60}$Co, and $^{137}$Cs were carefully evaluated for each batch, and the radioactivity 
	results with upper limits~\cite{Feldman_1998} at 90\% C.L. is shown in 
	Table~\ref{tab:summary_CLS_rate}.
	As the HPGe results are reported as upper limits, the batch with the best sensitivity --- batch \#3 --- was selected for installation, and its upper limits were used as the simulation inputs. This procedure significantly reduces the dominant background contribution from 
	the CLS cable.
	
	\begin{table}[htb]
		\centering
		\captionsetup{width=\textwidth}
		\caption{Example of material screening results for selected CLS cables batches.}
		\renewcommand{\arraystretch}{1.2}
		\begin{tabular}{c|c|c|c|c|c|c|c}
			\hline
			Component & Supplier & Unit &$^{238}$U & $^{232}$Th & $^{40}$K & $^{60}$Co & 
			$^{137}$Cs\\
			\hline
			\hline
			CLS cable \#1 & Fengshuo & mBq/kg & $< 57.11$ & $< 69.62$ & $< 649.64$ & 
			$< 11.04$ & $< 23.01$ \\
			CLS cable \#2 & Fengshuo & mBq/kg & $< 61.60$ & $< 109.26$ & $< 649.96$ 
			& $< 33.97$ & $< 49.32$ \\
			CLS cable \#3 & Fengshuo & mBq/kg & $< 43.76$ & $< 64.98$ & $< 334.46$ & 
			$< 45.35$ & $< 24.64$ \\
			CLS cable \#4 & Fengshuo & mBq/kg & $< 96.18$ & $< 73.14$ & $< 405.99$ & 
			$< 59.72$ & $< 66.32$ \\
			CLS cable \#5 & Fengshuo & mBq/kg & $< 141.11$ & $< 171.45$ & 
			$< 2105.78$ & $< 90.96$ & $< 62.92$ \\
			\hline
		\end{tabular}
		\label{tab:summary_CLS_rate}
	\end{table}
	
	The USS receivers were custom-developed for the JUNO calibration system with dedicated 
	low-background considerations. Different candidate materials and production batches were 
	systematically screened, and the components with the lowest radioactivity levels were selected 
	for assembly~\cite{Teng:2022usb}. The GTCS proximity sensors are mounted on the outer surface 
	of the acrylic vessel, where their $\gamma$ rays reach the liquid scintillator after traversing about 3~cm of water and the 12~cm acrylic wall, which attenuates them by approximately 50--70\% in the 1--3~MeV range (e.g. $\sim$60\% at 1.5~MeV and $\sim$50\% at 2.6~MeV), based on the NIST mass attenuation coefficients~\cite{NIST_XCOM}. Nevertheless, dedicated radiopurity screening was 
	carried out to further suppress the residual background contribution from these permanently 
	installed components. A similar screening strategy was applied to the GTCS proximity sensors. A 
	total of 40 sensors from different production batches were tested, and only those with 
	relatively low radioactive contamination were selected.
	
\begin{table*}[htbp]
	\centering
	\captionsetup{width=\textwidth}
	\caption{Radioactivity measurement results of the selected GTCS proximity sensors.}
    \tiny
	\label{tab:GTCS_sensor}
	\renewcommand{\arraystretch}{1.5}
	\begin{tabular}{c|c|c|c|c|c|c|c}
		\hline
		Component & Supplier & Unit & $^{238}$U & $^{232}$Th & $^{40}$K & $^{60}$Co & $^{137}$Cs \\
		\hline
		\hline
		Sensor \#6 & Omron & mBq/piece & $90.78 \pm 8.27$ & $169.12 \pm 15.24$ & $436.50 \pm 
		83.58$ & $< 9.46$ & $< 0.65$ \\
		Sensor \#9 & Omron & mBq/piece & $98.98 \pm 6.44$ & $176.00 \pm 13.48$ & $422.58 \pm 
		57.76$ & $< 2.35$ & $< 5.29$ \\
		Sensor \#12 & Omron & mBq/piece & $172.05 \pm 10.56$ & $188.26 \pm 14.33$ & $359.33 \pm 
		52.21$ & $< 0.48$ & $< 3.63$ \\
		Sensor \#21 & Omron & mBq/piece & $101.70 \pm 6.98$ & $205.83 \pm 15.77$ & $326.62 \pm 
		51.47$ & $< 2.71$ & $< 3.80$ \\
		Sensor \#22 & Omron & mBq/piece & $93.67 \pm 7.08$ & $192.68 \pm 15.39$ & $244.19 \pm 
		46.92$ & $< 1.56$ & $< 4.77$ \\
		Sensor \#23 & Omron & mBq/piece & $93.42 \pm 7.76$ & $190.52 \pm 16.30$ & $347.67 \pm 
		65.82$ & $< 3.50$ & $< 5.68$ \\
		Sensor \#25 & Omron & mBq/piece & $121.21 \pm 10.07$ & $180.00 \pm 15.97$ & $456.60 \pm 
		88.62$ & $< 0.72$ & $< 12.46$ \\
		Sensor \#26 & Omron & mBq/piece & $86.74 \pm 7.42$ & $190.72 \pm 15.71$ & $330.56 \pm 
		66.97$ & $< 4.97$ & $< 1.03$ \\
		Sensor \#29 & Omron & mBq/piece & $176.90 \pm 11.69$ & $180.51 \pm 14.54$ & $494.52 \pm 
		77.09$ & $< 1.57$ & $< 0.45$ \\
		Sensor \#31 & Omron & mBq/piece & $79.44 \pm 6.09$ & $166.26 \pm 13.19$ & $410.22 \pm 
		63.40$ & $< 1.23$ & $< 6.70$ \\
		\hline
	\end{tabular}
\end{table*}
	
    In addition, low-radioactivity materials such as PTFE and selected grades of stainless steel 
    were preferentially used for cables, support structures, and mechanical components. PTFE, known 
    for its high radiopurity, was extensively adopted for various calibration system components. 
    Its radioactivity was measured using NAA, which provides high 
    sensitivity for trace impurities. The PTFE material screened by NAA and that 
    used for the installed components are from the same manufacturer; a potential non-uniformity of 
    the trace radioactive contamination within the material may exist. These materials were 
    selected based on both assay results and 
    compatibility with detector requirements, ensuring a balance between mechanical performance and 
    radiopurity.
	
    The background associated with the calibration system constitutes a non-negligible part of the 
    overall detector background and must therefore be strictly controlled in JUNO. Compared with 
    typical commercial materials, the selected batches show significantly reduced radioactivity, 
    which is necessary to satisfy the stringent requirement of a total singles rate below 10 Hz in 
    the fiducial volume. This demonstrates that careful material screening is essential for 
    controlling calibration-induced background in JUNO. The final measured radioactivities of all 
    selected components are summarized in Table~\ref{tab:Radio_measure}.
	
\begin{figure}[!hb]
	\centering
	\includegraphics[width=3.0in]{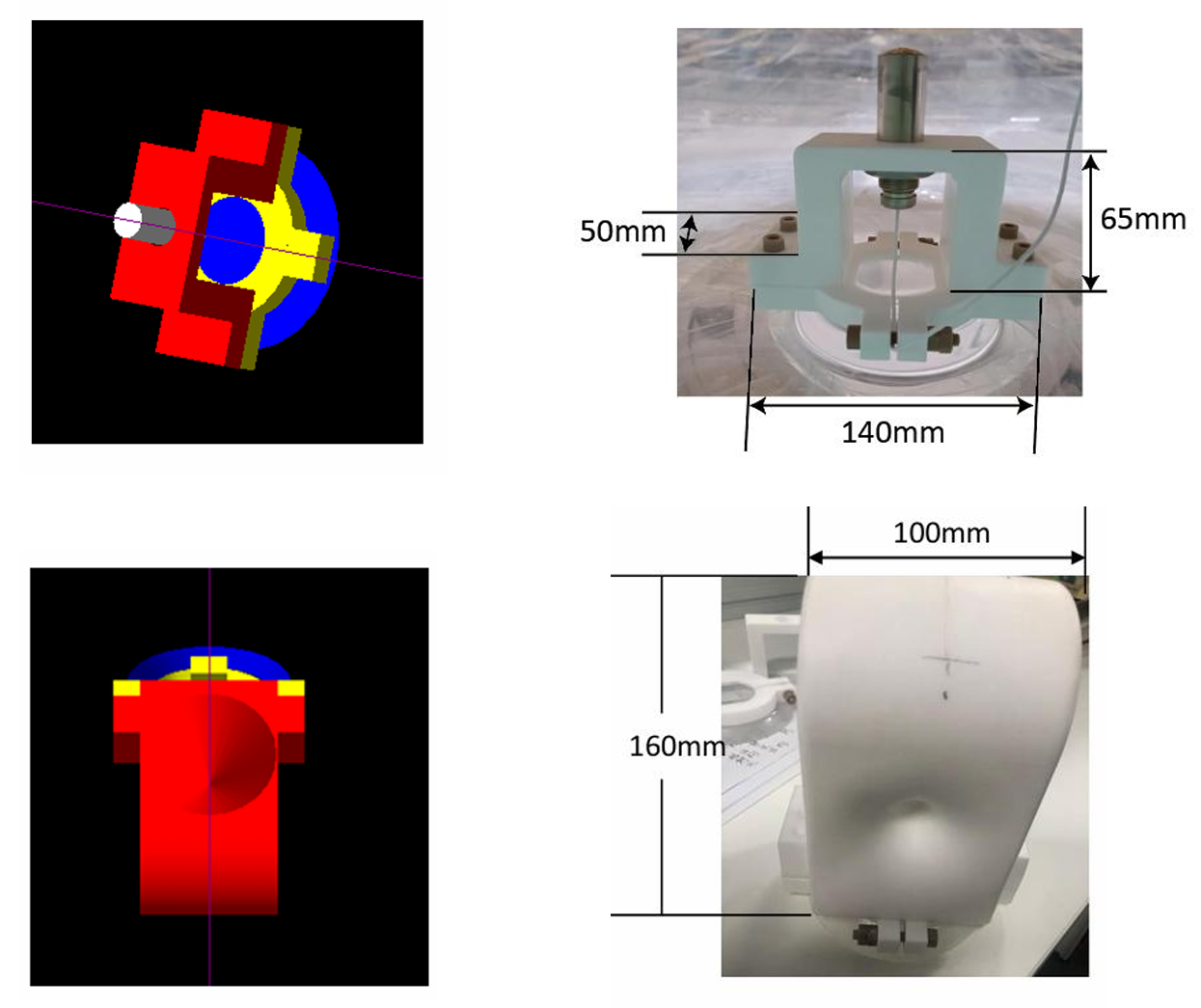}
	\caption{Geometry models in the simulation (left) and photograph of the actual components 
		(right) for selected calibration devices, including the ultrasonic receiver (top) and the 
		CLS cable anchor (bottom).}
	\label{fig:calibgeo}
\end{figure}%
	
	\section{Simulation of the calibration system background contribution}
	\label{sec:sim:cut}
	A Geant4-based detector simulation tool is developed for JUNO within the Software for 
	Non-collider Physics ExpeRiments (SNiPER) framework~\cite{geant4_paper,SNIPER}.
	Like Fig.~\ref{fig:calibgeo} shows, the calibration geometry for each component is implemented 
	into the SNiPER. 
	
    For each radioactive isotope ($^{238}\rm U$, $^{232}\rm Th$, $^{40}\rm K$, $^{60}\rm Co$, and 
    $^{137}$Cs), 
    events are generated uniformly within each component in the simulation. For the $^{238}\rm U$ 
    and $^{232}\rm Th$ decay chains, the early and late sub-chains are treated independently as discussed above. For the HPGe-based prediction, the late sub-chain upper limit is used as the chain-activity input, as the early sub-chain (tagged via the 186~keV $^{226}$Ra $\gamma$-ray) has low sensitivity. Since the radon migration process cannot be modeled reliably in the simulation, secular 
    equilibrium is assumed within each sub-chain when generating the Monte Carlo templates. For 
    each event, the deposited energy and its position in the LS are recorded in the simulation.
    
    \begin{figure}[!b]
	\centering
	\includegraphics[width=3.6in]{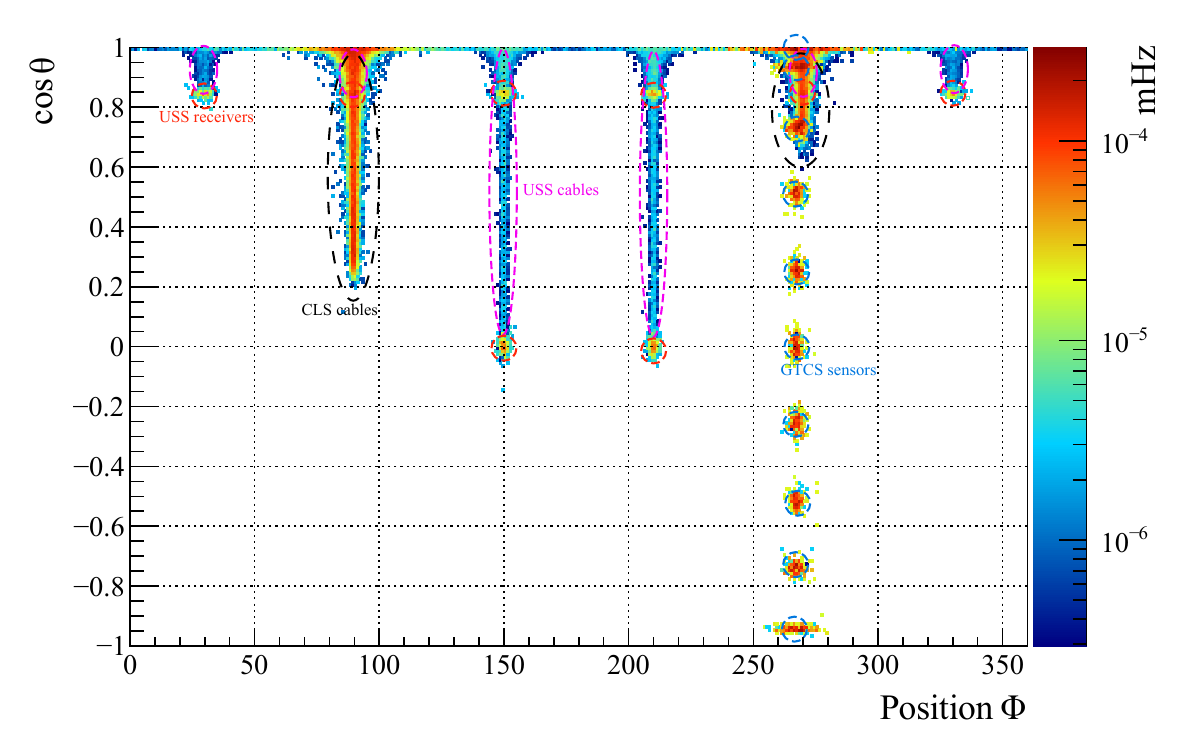}
	\caption{The overall spatial distribution of radioactivity from calibration system.
	}
	\label{calib:nocut:spectra}
\end{figure}
	 
\begin{figure}[!htbp]
 	\centering
 	\includegraphics[width=4.8in]{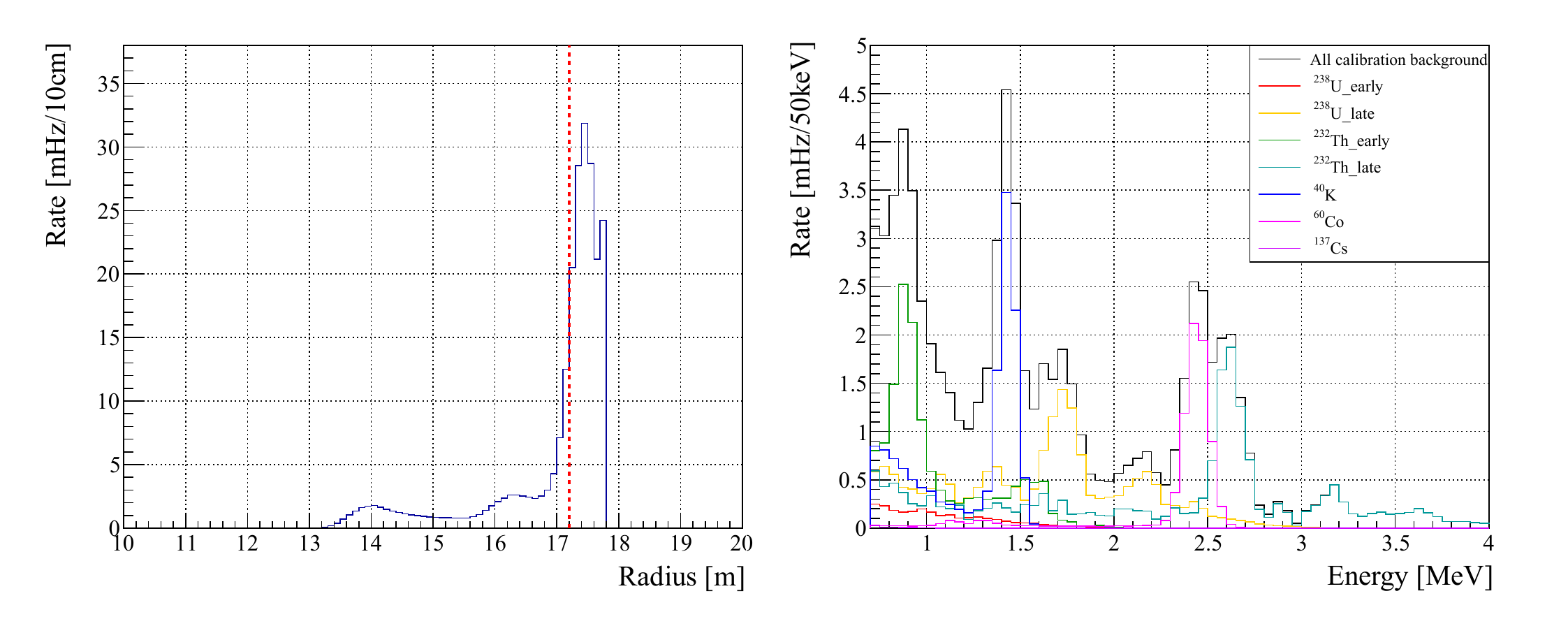}
 	\caption{
 		Left: Calibration radioactivity along the radius, the red line is at the 
 		fiducial volume cut with R $< 17.2$~m. 
 		Right: Simulated energy spectrum (after energy and fiducial volume selections) of the 
 		calibration-system background and its breakdown into individual radioactive components, 
 		including the early and late sub-chains of $^{238}\rm U$ and $^{232}\rm Th$, as well as 
 		$^{40}\rm K$, $^{60}\rm Co$ and $^{137}\rm Cs$. 
 		For the simulation, secular equilibrium is assumed within each decay sub-chain. 
 		Here "early" and "late" denote the sub-chains before and after radon 
 		($^{222}$Rn for $^{238}$U and $^{220}$Rn for $^{232}$Th).
 	}
 	\label{calib:energy:radius:cut}
 \end{figure}

    The detector response is modeled by applying an energy resolution of about $3\%/\sqrt{E}$ to 
    the deposited energy based on Ref.~\cite{CPC_energy_resolution}. The event vertex is smeared 
    with a position resolution of about $25~\mathrm{cm}/\sqrt{E}$~\cite{2025result1juno}. 
    This conservative value is adopted to bracket the worst-case vertex performance 
    near the detector edge; the typical resolution achieved by the standard reconstruction is about 
    15--20~cm. The choice is also well matched to the calibration-induced background, since most 
    calibration-system components (the calibration anchors, USS receivers, and GTCS sensors) are 
    located close to the acrylic sphere, near the detector periphery where the vertex resolution is 
    poorest; adopting 25~cm therefore provides a conservative estimate consistent with the spatial 
    distribution of these sources. For 
    reactor antineutrino detection, an energy threshold of $E > 0.7$~MeV is applied. Although a 
    fiducial volume cut with R $< 17.2$~m can effectively suppress 
    external background, the calibration components are not uniformly distributed within the JUNO 
    detector, as shown in Fig.~\ref{calib:nocut:spectra}. For CLS cables, they are at Y-Z plane, 
    and the shape is like catenary that the endpoints are at detector top chimney and detector edge 
    anchor. The USS receivers are at the equator and north latitude $58^{\circ}$ of acrylic inner 
    sphere. The components of the calibration system can be modeled as catenaries (e.g., the CLS 
    cable) or localized point sources (e.g., USS receiver with 6~cm length) compared with the 
    35.4~m diameter LS detector. Furthermore, a radius cut is not entirely suitable for the CLS 
    cables. With energy cut E $> 0.7$~MeV, Fig.~\ref{calib:energy:radius:cut} shows 
    the calibration background distribution along radius. After applying the fiducial volume cut, about $37\%$ of the calibration-induced background survives and is to be estimated.
    
	The survival probability of each isotope is evaluated using the simulation with energy 
	and fiducial volume selections. Based on the material radioactivity (shown in 
	Table~\ref{tab:Radio_measure}) measured by HPGe 
	detectors/NAA and the simulated survival probability, the expected calibration system 
	background rate can be estimated as
	\begin{equation}
		R = A \times P,
	\end{equation}
	where $A$ is the activity of the radioactive isotope in each component and $P$ is the 
	corresponding survival probability after all selections. The total contribution from the 
	calibration system is estimated to be less than 76~mHz. Since some of the material 
	radioactivity measurements are reported as upper limits, these upper limits are used directly as the activity inputs, yielding a conservative overestimate of the background. A detailed breakdown of the background contributions from 
	different subsystems is summarized in Table~\ref{tab:summary_rate_radius_cut}.
	The cuts are not fully optimized for the non-spherically symmetric calibration system. 
	Nevertheless, the residual background contribution after applying standard selections is 
	sufficiently small, indicating that no dedicated calibration-specific fiducial volume cut 
	is required for the physics analysis.
		
\begin{table}[htbp]
	\centering
    \captionsetup{width=\textwidth}  
    \caption{Summary of single rates from the calibration system.}
	\renewcommand{\arraystretch}{1.2}
	\begin{tabular}{c|c|c|c}  \hline System & Item & Quantity & Rate (mHz) \\ 
	\hline \hline \multirow{2}{*}{CLS} &   cables & 0.27 kg & $< 38.35$  \\ &   
	anchors  & 6.30 kg & $< 0.04$ \\ \hline \multirow{3}{*}{USS} & receivers  & 
	10 pieces & $< 3.98$  \\ & cables     & 1.45 kg & $< 3.37$  
	\\ 
	& anchors  & 5.56 kg & $< 0.03$ \\ \hline \multirow{1}{*}{GTCS} & sensors  & 
	10 pieces & $< 30.18$ \\ \hline Total & -- & -- & $< 75.95$ 
	\\ 
	\hline \end{tabular}
	\label{tab:summary_rate_radius_cut}  
\end{table}

	Accidental background is one of the important backgrounds in the reactor antineutrino analysis 
	of JUNO. IBD events are identified through the characteristic coincidence of a prompt positron 
	signal and a delayed neutron-capture signal within a time window of approximately 1000~$\mu$s. 
	Accidental background originates primarily from random pairs of uncorrelated radioactive 
	decays~\cite{JUNO-bkg-paper}. When two independent radioactivity-induced events accidentally 
	satisfy all IBD selection criteria, they can mimic IBD candidates and consequently contaminate 
	the neutrino signal sample. Therefore, a reliable estimation of the accidental background rate 
	is essential for evaluating the influence of background contributions introduced by the 
	calibration system. The accidental background rate is defined as
	
	\begin{equation}
		R_{\text{acc}} = R_{p} \times R_{d} \times T \times \epsilon_{r},
	\end{equation}
	where $R_{p}$ and $R_{d}$ denote the prompt-like and delayed-like singles rates, respectively, 
	$T$ is the prompt-delayed coincidence time window, and $\epsilon_{r}$ represents the efficiency 
	for two uncorrelated signals to pass the spatial distance cut, which is set to 1.5~m in this 
	study.
	
	The potential delayed-like signals mainly include radioactive decays, cosmogenic isotopes, and 
	spallation neutrons. However, compared with intrinsic radioactivity, the contributions from 
	cosmogenic isotopes and spallation neutrons to the delayed-like background are sufficiently 
	small after applying the muon veto, which vetoes the entire detector for 5~ms following each detected muon~\cite{2025result1juno}, and can be safely neglected in the present analysis. Consequently, only 
	radioactivity-induced signals are considered for the estimation of the accidental background 
	related to the calibration system.
	
	In the simulation, the target impurity concentrations of different detector materials follow 
	the design values specified in the JUNO background control strategy paper~\cite{yellow-book}. 
	Within the fiducial	volume defined by $r < 17.2$~m and the energy range from 0.7~MeV to 
	12~MeV, the total singles rate from detector is about 7.2~Hz. Among these 
	events, 
	approximately 15\% fall into the delayed-like energy window between 1.9~MeV and 2.5~MeV. The 
	time coincidence window is 
	chosen to be $T = 1$~ms, consistent with the standard IBD selection criteria adopted in JUNO. 
	The increase in the accidental background rate induced by the calibration system can be 
	expressed as
	\begin{multline}
		\delta R = (R_{p}^{0} + R_{p}^{\text{calib}})\times(R_{d}^{0} + R_{d}^{\text{calib}})\times 
		T\times\epsilon_{r}^{\text{calib}} - R_{p}^{0} \times R_{d}^{0} \times 
		T\times\epsilon_{r}^{0},
	\end{multline}
	where $R_{p}^{0}$ and $R_{d}^{0}$ are the prompt-like and delayed-like radioactivity rates in 
	the absence of the calibration system, while $R_{p}^{\text{calib}}$ and $R_{d}^{\text{calib}}$ 
	denote the corresponding rates introduced solely by the calibration system components. The 
	parameters $\epsilon_{r}^{0}$ and $\epsilon_{r}^{\text{calib}}$ represent the spatial 
	coincidence efficiencies without and with the calibration system, respectively.
	
	After applying a prompt-delayed spatial distance cut of less than 1.5~m, a suppression factor 
	of approximately 300 is achieved for uncorrelated accidental pairs. Taking this spatial cut 
	into account, the resulting increase in the accidental background rate due to the calibration 
	system is estimated to be about $1\times10^{-3}$ events per day. This contribution is 
	negligibly small compared with the expected IBD signal rate of about 48 events per day in JUNO 
	with efficiency correction. 
	Therefore, the additional accidental background introduced by the calibration system has an 
	insignificant impact on the reactor antineutrino measurements, and no dedicated 
	calibration-specific fiducial volume cut is required in the standard IBD event selection.

\section{In-situ radioactive background results for calibration system}

The JUNO detector started physics data taking in August 2025. To evaluate the in-situ radioactive 
background contribution from the calibration system, we use the first 59~days of data collected 
since the detector completion. This period corresponds to stable detector operating conditions with 
the calibration system parked in its standard non-deployment position. 
All events were reconstructed using the standard JUNO reconstruction chain (VTREP for both energy 
and position reconstruction~\cite{Takenaka_2025}), and data-quality 
selections were applied to remove instrumental noise, unstable electronics periods, 
and the standard muon veto was also applied.

Simulation studies indicate that the dominant calibration-related contributions inside the 
liquid scintillator originate from the CLS cables, USS receivers, and GTCS proximity sensors. 
These contributions exhibit distinct spatial features, including an extended structure associated 
with the CLS cables and localized features related to the USS receivers and GTCS sensors, which are 
analyzed separately in the following.
\subsection{Spatial distribution along the CLS cable}

\begin{figure}[H]
	\centering
	\includegraphics[width=3.2in]{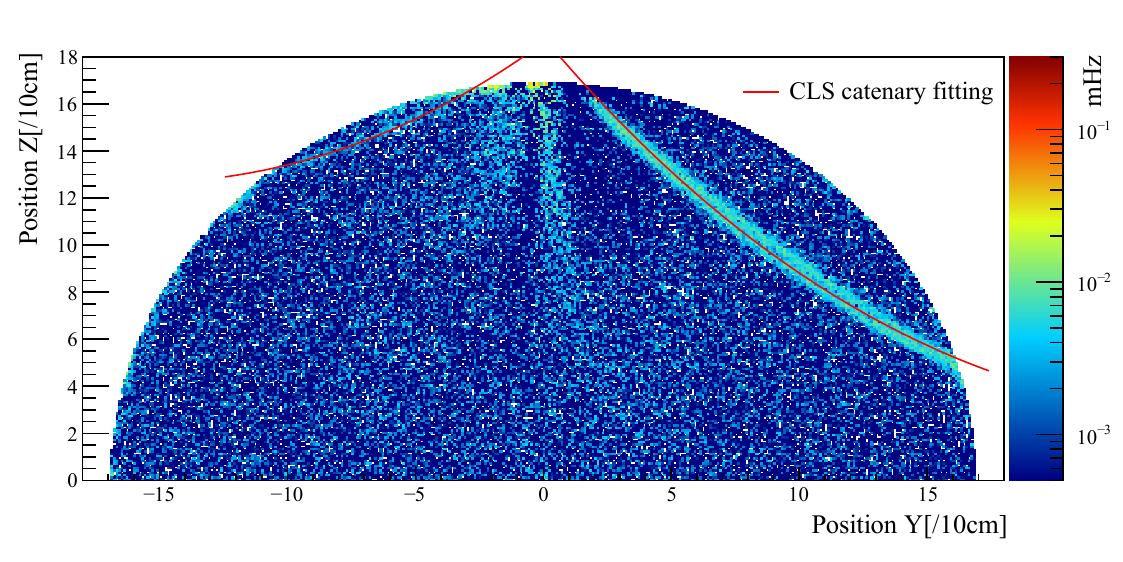}
	\caption{
		The $Z$--$Y$ plane distribution of CLS cable-related events.
	}
	\label{calib:CLScut:2D}
\end{figure}

The CLS cable forms a suspended catenary structure, leading to an extended spatial distribution 
of $\gamma$-induced events. The structure models the equilibrium shape of a flexible cable 
suspended under gravity. 
In this coordinate system, the vertical direction corresponds to the $Z$ axis, 
while $Y$ represents the transverse coordinate in the plane. The cable trajectory is modeled as

\begin{equation}
	z(y) = a \cosh\left(\frac{y - y_0}{a}\right) + z_0,
\end{equation}

where $a$ characterizes the curvature of the cable, and $(y_0, z_0)$ defines its position. 
This function provides a good approximation of the cable geometry and is used to fit the 
reconstructed event distribution associated with the CLS.
Fig.~\ref{calib:CLScut:2D} shows the reconstructed event distribution in the $Z$--$Y$ plane after 
applying a fiducial volume cut of $R<16.5$~m for better visualization and an energy threshold of 
$E>0.7$~MeV. The corresponding feature is clearly observed in data and is consistent with 
simulation. This spatial characteristic provides an effective handle for identifying events 
associated with the CLS cable and serves as the basis for the quantitative background extraction 
described in the following section.

\begin{figure}[!htbp]
	\centering
	\includegraphics[width=4.8in]{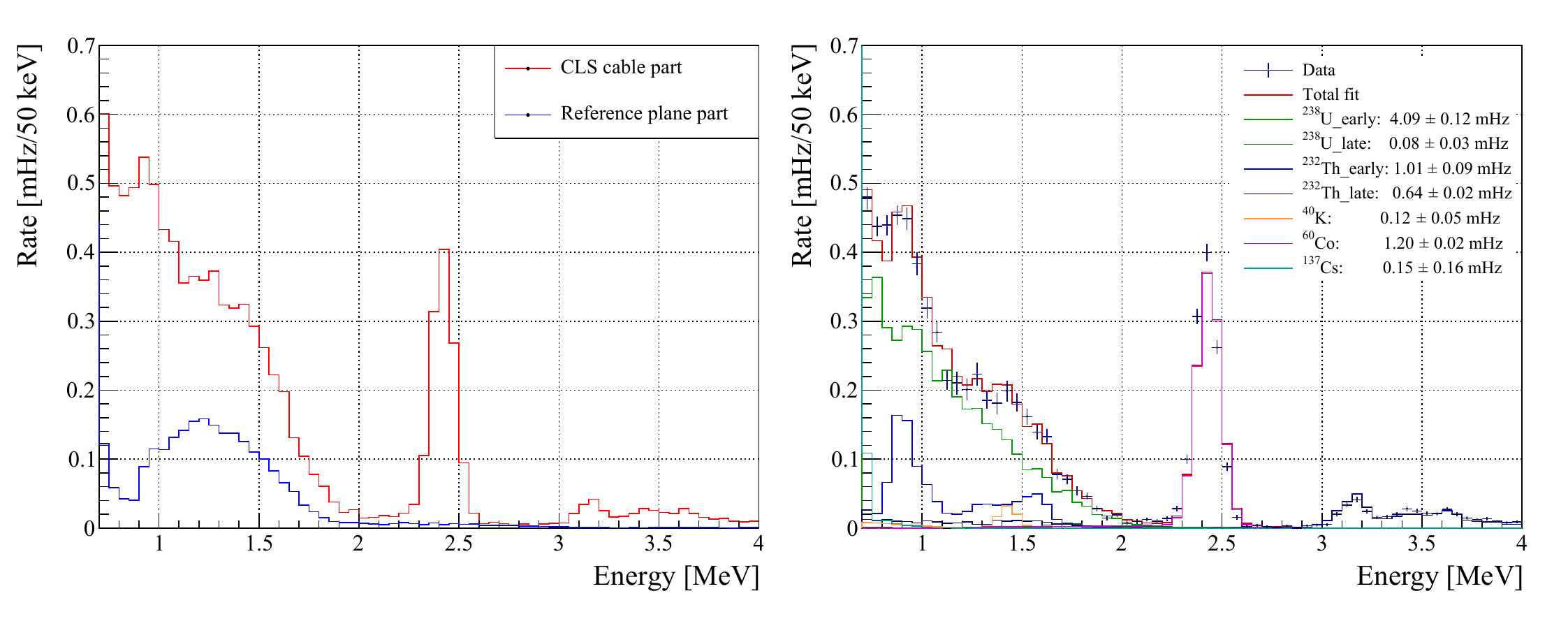}
	\caption{
		Left: The tagged CLS cable radioactivity spectrum(red) and reference plane spectrum(blue).
		Right: The reference-subtracted CLS spectrum and breakdown fitting for each radioactive 
		isotope.
	}
	\label{calib:CLS:cut}
\end{figure}

To isolate the CLS-related contribution, the cable track is reconstructed using the catenary function of Eq.~(4). A cylindrical region with a radius of 1~m around the fitted curve is defined as the 
signal region. In addition, a vertical sheet with $X\in[-1,1]$~m is selected to contain the CLS 
trajectory. After the calibration source is retracted, the side cable remains as a double steel strand; the simulation models both strands with their actual length and a separation of 20~cm, while a single catenary is used for the event selection since the two strands are not resolved at the present position resolution. The natural radioactive background is estimated using several reference planes located 
at $\phi=45^\circ$, $90^\circ$, $135^\circ$, $225^\circ$, $270^\circ$, and $315^\circ$. The same 
geometrical selection criteria are applied in each reference plane, and the resulting event samples 
are averaged to obtain the background expectation. After normalization, the averaged background 
spectrum is subtracted from the signal-region sample to remove contributions from bulk 
radioactivity. The remaining excess of events associated with the cable trajectory is attributed to 
the radioactivity of the CLS cable. The corresponding energy spectrum is shown in 
Fig.~\ref{calib:CLS:cut}.

The residual spectrum is further analyzed using a template fit based on the simulated energy 
spectra of different radioactive components, including the early and late sub-chains of $^{238}\rm 
U$ and $^{232}\rm Th$, as well as $^{40}\rm K$, $^{60}\rm Co$ and $^{137}\rm Cs$. The activities 
associated with the CLS cable are extracted from the spectral fit. The fitted results are compared 
with the expectations derived from HPGe material assay combined with Monte Carlo simulation. For 
the HPGe-based prediction, the $^{238}\rm U$ and $^{232}\rm Th$ activities are evaluated as total 
upper limits without separating the early and late sub-chains, resulting in conservative estimates 
for the corresponding components. A quantitative comparison of the fitted activities and the 
corresponding predictions is summarized in Table~\ref{tab:cls_activity_compare}. Overall, the 
data-driven results are found to be consistent with these conservative predictions, indicating that 
both the material radioactivity inputs and the detector response modeling are reliable.

\begin{table}[ht]
	\centering
	\captionsetup{width=\textwidth}
	\caption{Comparison of radioactive activities associated with the one set of CLS cable 
		obtained from data fitting and from HPGe/NAA measurement combined with Monte Carlo 
		simulation. For the U/Th in the data fit, the early and late sub-chain 
		results are combined into a single chain activity ($A = A_{\rm early} + A_{\rm late}$), 
		with the uncertainty propagated in quadrature.}
	\renewcommand{\arraystretch}{1.2}
	\label{tab:cls_activity_compare}
	\begin{tabular}{c|c|c}
		\hline
		Isotope / Component & Data (fit) [mHz] & HPGe + MC [mHz] \\
		\hline
		\hline
		$^{238}\rm U$           & $4.17 \pm 0.12$ & $< 8.23$ \\
		\hline
		$^{232}\rm Th$          & $1.65 \pm 0.09$ & $< 7.34$ \\
		\hline
		$^{40}\rm K$           & $0.12 \pm 0.05$ & $< 5.07$ \\
		\hline
		$^{60}\rm Co$          & $1.20 \pm 0.02$ & $< 4.51$ \\
		\hline
		$^{137}\rm Cs$         & $0.15 \pm 0.16$ & $< 0.04$ \\
		\hline
		Total                 & $7.29$ & $< 25.19$ \\
		\hline
	\end{tabular}
\end{table}

\FloatBarrier
\subsection{Localized spatial distributions: USS and GTCS}

\begin{figure}[h]
	\centering
 	\includegraphics[width=3.5in]{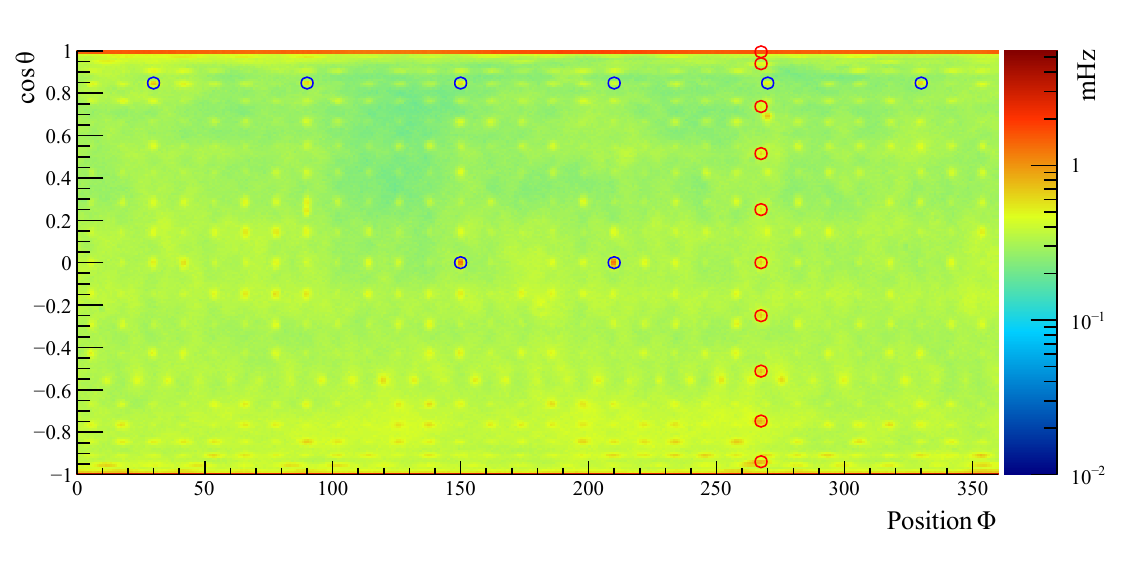}
	\caption{
		The $\phi$--cos $\theta$ plane distribution of USS receiver-like (blue) and GTCS 
		sensor-like 
		(red) 
		events.
		A fiducial volume cut of $R<17.2$~m and an energy cut of $E > 0.7$~MeV are applied. 
		The additional hot spots visible in the figure are associated with other 
		detector structures (e.g., the acrylic sphere nodes); the blue and red circles mark the 
		USS-receiver and GTCS-sensor positions, respectively.
	}
	\label{calib:USScut:2D}
\end{figure}

The USS receivers and GTCS proximity sensors appear as hot spots in Fig.~\ref{calib:USScut:2D}. 
The USS receivers are attached to the inner surface of the acrylic vessel, while the GTCS sensors 
are distributed on the outer surface at a fixed azimuthal angle and varying polar angles, forming 
several localized features in detector coordinates.

To evaluate their contributions, cylindrical selection regions are defined around each USS receiver 
and sensor position. The cylinder is constructed along the radial direction of the detector with a 
radius of 0.5~m. Only the part within the radial range of $16.5<R<17.2$~m is used in the analysis.
This geometry is motivated by the fact that the excess events associated with the USS receivers and 
GTCS sensors are concentrated around the sensor positions and extend approximately along the radial 
direction into the liquid scintillator.
The selection criteria are determined in a data-driven manner. The selection range is varied to
ensure that the excess associated with the sensors is fully captured, while the event rate in the inner region ($R<16.5$~m) remains consistent with the LS background estimated from a reference location
at the same radius. 
The reference background is estimated from a location at $\phi=100^\circ$ and $\theta=90^\circ$, 
where no calibration-system components are present. An identical geometrical selection is applied 
to the reference location and the sensor positions. After normalization, the reference sample is 
subtracted from the sensor-associated sample to isolate the excess events associated with the USS 
receivers and GTCS sensors.
The residual spectrum exhibits the characteristic $^{60}$Co $\gamma$ peak, which is fitted to
confirm the presence of sensor-related radioactivity and to estimate its contribution. The overall
background contribution is subsequently evaluated using the total singles rate after background
subtraction.

\begin{figure}[!htbp]
	\centering
	\includegraphics[width=4.8in]{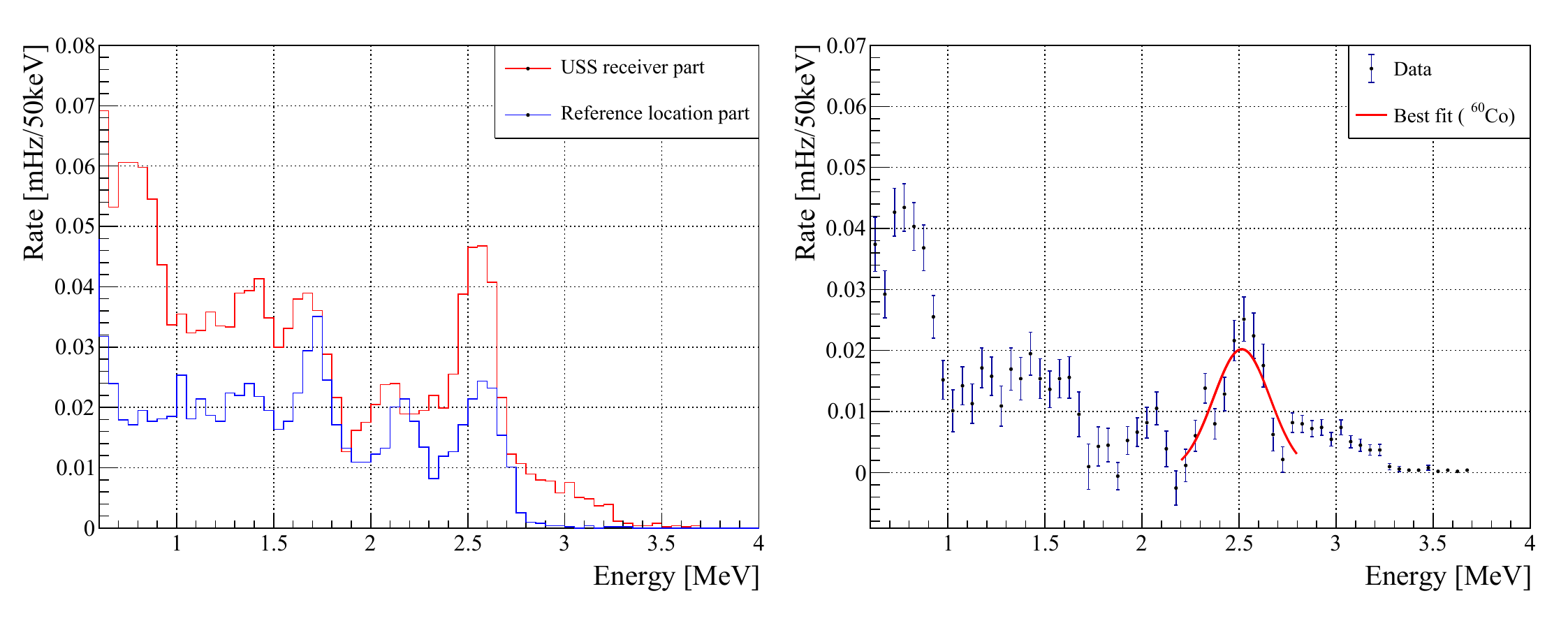}
	\caption{
		Left: The tagged USS receiver-related radioactivity spectrum(red) and reference location 
		spectrum(blue).
		Right: The background-subtracted USS receiver spectrum.
	}
	\label{calib:USS:cut}
\end{figure}

\begin{figure}[!htbp]
	\centering
	\includegraphics[width=4.8in]{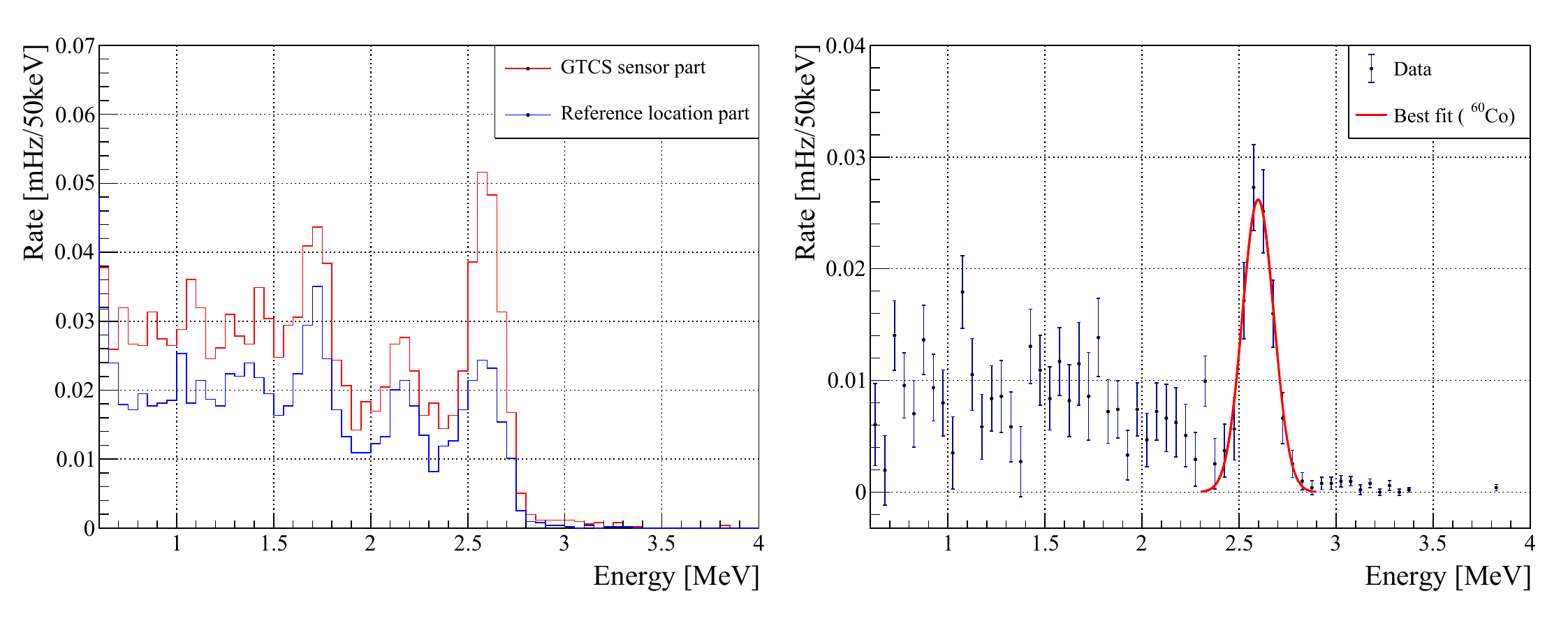}
	\caption{
		Left: The tagged GTCS sensor-related radioactivity spectrum(red) and reference location 
		spectrum(blue).
		Right: The background-subtracted GTCS sensor spectrum.
	}
	\label{calib:GTCS:cut}
\end{figure}

The background contributions associated with the USS receivers and GTCS sensors are evaluated 
using the data-driven subtraction method described above. And the spectra are shown as 
Fig.~\ref{calib:USS:cut} and Fig.~\ref{calib:GTCS:cut}. The extracted rates are found to be 
$(3.24 \pm 0.10)$~mHz for the USS receivers and $(3.93 \pm 0.19)$~mHz for the GTCS sensors.
These values are consistent with the expectations derived from the HPGe measurements and Monte 
Carlo simulations, which predict contributions below 4.01~mHz and 30.18~mHz, respectively. 

The results obtained from the in-situ data analysis are consistent with the expectations based on 
material radioactivity measurements and detector simulation. 
In particular, the background contribution associated with the CLS cable is found to be of the same 
order of magnitude as the prediction derived from HPGe assay and Monte Carlo simulation, 
indicating that the geometrical modeling and efficiency estimation are reliable.
For the GTCS proximity sensors, the background contribution observed in data is lower than the 
expectation based on the measured material activities. Since the HPGe results are typically 
reported as upper limits, the corresponding predictions are conservative. The observed difference 
is therefore consistent within uncertainties.

Overall, the agreement between data and prediction demonstrates that the background contributions 
from the calibration system are well understood and within prediction. 
After applying all physics selection criteria, the residual background contribution from 
the entire calibration system, including the CLS cables, the USS receivers, and the GTCS proximity 
sensors, corresponds to less than 2\% of the total background rate in the reactor antineutrino 
energy window of 0.7--12~MeV. 
From the perspective of physics analysis, this contribution is negligible compared with the IBD 
signal rate and does not introduce observable distortion in the reconstructed energy spectrum.

	\section{Conclusion}
	With the HPGe detector and NAA assay, we have measured the radioactivity of all the materials 
	of calibration system that potentially introduce background, and selected the lowest 
	radioactivity from several candidate materials. Furthermore, utilizing a Geant4-based detector 
	simulation,	we have calculated background efficiency of every component from calibration system.
	The single rate of background from calibration system is no more than 76~mHz, 
	which meets the requirement($<$ 200 mHz). In addition, the increased rate of 
	coincidence IBD background due to calibration is 0.001 per day, which is a very small value 
	compared with IBD rate of about 48 per day with efficiency correction~\cite{2026result1juno}. The in-situ detector data confirm these expectations, with the CLS-cable, USS-receiver, and GTCS-sensor backgrounds consistent with the HPGe-assay-plus-MC predictions within uncertainties.
	
	These results demonstrate that the low-background design and material control of the JUNO 
	calibration system are effective, and that its background contribution to reactor antineutrino 
	detection can be neglected under standard IBD selection criteria. The methodology presented in 
	this work also provides a useful reference for the background assessment and material 
	optimization of calibration devices in future low-background neutrino experiments.

    \section*{Acknowledgments}
    The authors gratefully acknowledge the support from the PandaX collaboration 
    for the radioactivity measurements of the calibration-system materials with the HPGe in CJPL. 
    We sincerely thank Monica Sisti and collaborators at INFN Milano Bicocca for the neutron 
    activation analysis (NAA) measurements of PTFE samples. We also appreciate the valuable 
    discussions and support from the JUNO collaborators.
    
    This work is supported by the National Key Research and Development Program of China(Grant no. 
    2023YFA1606104),  National Science Foundation of China (Grant number: 12222505), and the 
    Strategic Priority Research Program of the Chinese Academy of Sciences(Grant number: 
    XDA10010800), Y. M. and R. L. thank the sponsorship from the Yangyang Development Fund.
    \bibliography{references}


\begin{thebibliography}{22}
\ifx \bisbn   \undefined \def \bisbn  #1{ISBN #1}\fi
\ifx \binits  \undefined \def \binits#1{#1}\fi
\ifx \bauthor  \undefined \def \bauthor#1{#1}\fi
\ifx \batitle  \undefined \def \batitle#1{#1}\fi
\ifx \bjtitle  \undefined \def \bjtitle#1{#1}\fi
\ifx \bvolume  \undefined \def \bvolume#1{\textbf{#1}}\fi
\ifx \byear  \undefined \def \byear#1{#1}\fi
\ifx \bissue  \undefined \def \bissue#1{#1}\fi
\ifx \bfpage  \undefined \def \bfpage#1{#1}\fi
\ifx \blpage  \undefined \def \blpage #1{#1}\fi
\ifx \burl  \undefined \def \burl#1{\textsf{#1}}\fi
\ifx \doiurl  \undefined \def \doiurl#1{\url{https://doi.org/#1}}\fi
\ifx \betal  \undefined \def \betal{\textit{et al.}}\fi
\ifx \binstitute  \undefined \def \binstitute#1{#1}\fi
\ifx \binstitutionaled  \undefined \def \binstitutionaled#1{#1}\fi
\ifx \bctitle  \undefined \def \bctitle#1{#1}\fi
\ifx \beditor  \undefined \def \beditor#1{#1}\fi
\ifx \bpublisher  \undefined \def \bpublisher#1{#1}\fi
\ifx \bbtitle  \undefined \def \bbtitle#1{#1}\fi
\ifx \bedition  \undefined \def \bedition#1{#1}\fi
\ifx \bseriesno  \undefined \def \bseriesno#1{#1}\fi
\ifx \blocation  \undefined \def \blocation#1{#1}\fi
\ifx \bsertitle  \undefined \def \bsertitle#1{#1}\fi
\ifx \bsnm \undefined \def \bsnm#1{#1}\fi
\ifx \bsuffix \undefined \def \bsuffix#1{#1}\fi
\ifx \bparticle \undefined \def \bparticle#1{#1}\fi
\ifx \barticle \undefined \def \barticle#1{#1}\fi
\bibcommenthead
\ifx \bconfdate \undefined \def \bconfdate #1{#1}\fi
\ifx \botherref \undefined \def \botherref #1{#1}\fi
\ifx \url \undefined \def \url#1{\textsf{#1}}\fi
\ifx \bchapter \undefined \def \bchapter#1{#1}\fi
\ifx \bbook \undefined \def \bbook#1{#1}\fi
\ifx \bcomment \undefined \def \bcomment#1{#1}\fi
\ifx \oauthor \undefined \def \oauthor#1{#1}\fi
\ifx \citeauthoryear \undefined \def \citeauthoryear#1{#1}\fi
\ifx \endbibitem  \undefined \def \endbibitem {}\fi
\ifx \bconflocation  \undefined \def \bconflocation#1{#1}\fi
\ifx \arxivurl  \undefined \def \arxivurl#1{\textsf{#1}}\fi
\csname PreBibitemsHook\endcsname

\bibitem[\protect\citeauthoryear{An et~al.}{2016}]{yellow-book}
\begin{barticle}
\bauthor{\bsnm{An}, \binits{F.}}, \betal:
\batitle{Neutrino physics with {JUNO}}.
\bjtitle{Journal of Physics G: Nuclear and Particle Physics}
\bvolume{43}(\bissue{3}),
\bfpage{030401}
(\byear{2016})
\doiurl{10.1088/0954-3899/43/3/030401}
\end{barticle}
\endbibitem

\bibitem[\protect\citeauthoryear{}{2022}]{PPNP}
\begin{botherref}
{JUNO} physics and detector.
Progress in Particle and Nuclear Physics
\textbf{123},
103927
(2022)
\doiurl{10.1016/j.ppnp.2021.103927}
\end{botherref}
\endbibitem

\bibitem[\protect\citeauthoryear{Sisti and Zhao}{2020}]{Sisti2020LowBackground}
\begin{botherref}
\oauthor{\bsnm{Sisti}, \binits{M.}},
\oauthor{\bsnm{Zhao}, \binits{J.}}:
{LOW BACKGROUND Task Force}.
Report,
JUNO Collaboration
(2020).
Presented at the 2nd JUNO International Scientific Committee Meeting
\end{botherref}
\endbibitem

\bibitem[\protect\citeauthoryear{and Abusleme
  et~al.}{2021}]{calib_strategy_paper}
\begin{botherref}
\oauthor{\bsnm{Abusleme}, \binits{A.}},
\oauthor{\bsnm{Adam}, \binits{T.}},
\oauthor{\bsnm{Ahmad}, \binits{S.}}, et al.:
Calibration strategy of the {JUNO} experiment.
Journal of High Energy Physics
\textbf{2021}(3)
(2021)
\doiurl{10.1007/jhep03(2021)004}
\end{botherref}
\endbibitem

\bibitem[\protect\citeauthoryear{Hui et~al.}{2021}]{JUNO-ACU-paper}
\begin{barticle}
\bauthor{\bsnm{Hui}, \binits{J.}},
\bauthor{\bsnm{Liu}, \binits{H.}},
\bauthor{\bsnm{Liu}, \binits{J.}},
\bauthor{\bsnm{Meng}, \binits{Y.}},
\bauthor{\bsnm{Xiao}, \binits{M.}},
\bauthor{\bsnm{Xu}, \binits{D.}},
\bauthor{\bsnm{Yang}, \binits{L.}},
\bauthor{\bsnm{Ye}, \binits{Z.}},
\bauthor{\bsnm{Zhang}, \binits{F.}},
\bauthor{\bsnm{Zhang}, \binits{T.}},
\bauthor{\bsnm{Zhang}, \binits{Y.}}:
\batitle{The automatic calibration unit in {JUNO}}.
\bjtitle{Journal of Instrumentation}
\bvolume{16}(\bissue{08}),
\bfpage{08008}
(\byear{2021})
\doiurl{10.1088/1748-0221/16/08/t08008}
\end{barticle}
\endbibitem

\bibitem[\protect\citeauthoryear{Zhang et~al.}{2019}]{Zhang:2018yso}
\begin{barticle}
\bauthor{\bsnm{Zhang}, \binits{Y.}},
\bauthor{\bsnm{Liu}, \binits{J.}},
\bauthor{\bsnm{Xiao}, \binits{M.}},
\bauthor{\bsnm{Zhang}, \binits{F.}},
\bauthor{\bsnm{Zhang}, \binits{T.}}:
\batitle{Laser calibration system in {JUNO}}.
\bjtitle{Journal of Instrumentation}
\bvolume{14}(\bissue{01}),
\bfpage{01009}--\blpage{01009}
(\byear{2019})
\doiurl{10.1088/1748-0221/14/01/p01009}
\end{barticle}
\endbibitem

\bibitem[\protect\citeauthoryear{Takenaka et~al.}{2024}]{Takenaka:2024ctk}
\begin{barticle}
\bauthor{\bsnm{Takenaka}, \binits{A.}},
\bauthor{\bsnm{Hui}, \binits{J.}},
\bauthor{\bsnm{Li}, \binits{R.}},
\bauthor{\bsnm{Hao}, \binits{S.}},
\bauthor{\bsnm{Huang}, \binits{J.}},
\bauthor{\bsnm{Lai}, \binits{H.}},
\bauthor{\bsnm{Li}, \binits{Y.}},
\bauthor{\bsnm{Liu}, \binits{J.}},
\bauthor{\bsnm{Meng}, \binits{Y.}},
\bauthor{\bsnm{Qian}, \binits{Z.}},
\bauthor{\bsnm{Wang}, \binits{H.}},
\bauthor{\bsnm{Xiang}, \binits{Z.}},
\bauthor{\bsnm{Yuan}, \binits{Z.}},
\bauthor{\bsnm{Yun}, \binits{Y.}},
\bauthor{\bsnm{Zhang}, \binits{F.}},
\bauthor{\bsnm{Zhang}, \binits{T.}},
\bauthor{\bsnm{Zhang}, \binits{Y.}}:
\batitle{Customized calibration sources in the {JUNO} experiment}.
\bjtitle{Journal of Instrumentation}
\bvolume{19}(\bissue{12}),
\bfpage{12019}
(\byear{2024})
\doiurl{10.1088/1748-0221/19/12/p12019}
\end{barticle}
\endbibitem

\bibitem[\protect\citeauthoryear{Zhang et~al.}{2021}]{zhangyy-paper}
\begin{barticle}
\bauthor{\bsnm{Zhang}, \binits{Y.}},
\bauthor{\bsnm{Hui}, \binits{J.}},
\bauthor{\bsnm{Liu}, \binits{J.}},
\bauthor{\bsnm{Xiao}, \binits{M.}},
\bauthor{\bsnm{Zhang}, \binits{T.}},
\bauthor{\bsnm{Zhang}, \binits{F.}},
\bauthor{\bsnm{Meng}, \binits{Y.}},
\bauthor{\bsnm{Xu}, \binits{D.}},
\bauthor{\bsnm{Ye}, \binits{Z.}}:
\batitle{{Cable Loop Calibration System for Jiangmen Underground Neutrino
  Observatory}}.
\bjtitle{Nucl. Instrum. Meth. A}
\bvolume{988},
\bfpage{164867}
(\byear{2021})
\doiurl{10.1016/j.nima.2020.164867}
{\href{https://arxiv.org/abs/2011.02183}{{arXiv:2011.02183}}}
{[physics.ins-det]}
\end{barticle}
\endbibitem

\bibitem[\protect\citeauthoryear{Hui et~al.}{2025}]{Hui_2025}
\begin{barticle}
\bauthor{\bsnm{Hui}, \binits{J.}},
\bauthor{\bsnm{Li}, \binits{R.}},
\bauthor{\bsnm{Wu}, \binits{Y.}},
\bauthor{\bsnm{Zhang}, \binits{T.}},
\bauthor{\bsnm{Chen}, \binits{Z.}},
\bauthor{\bsnm{Freegard}, \binits{A.}},
\bauthor{\bsnm{Huang}, \binits{J.}},
\bauthor{\bsnm{Lai}, \binits{H.}},
\bauthor{\bsnm{Liao}, \binits{Y.}},
\bauthor{\bsnm{Liu}, \binits{J.}},
\bauthor{\bsnm{Meng}, \binits{Y.}},
\bauthor{\bsnm{Takenaka}, \binits{A.}},
\bauthor{\bsnm{Xiang}, \binits{Z.}},
\bauthor{\bsnm{Zhang}, \binits{P.}},
\bauthor{\bsnm{Zhang}, \binits{Y.}}:
\batitle{The calibration house in {JUNO}}.
\bjtitle{Journal of Instrumentation}
\bvolume{20}(\bissue{10}),
\bfpage{10035}
(\byear{2025})
\doiurl{10.1088/1748-0221/20/10/P10035}
\end{barticle}
\endbibitem

\bibitem[\protect\citeauthoryear{Guo et~al.}{2019}]{GTCS}
\begin{barticle}
\bauthor{\bsnm{Guo}, \binits{Y.}},
\bauthor{\bsnm{Zhang}, \binits{Q.}},
\bauthor{\bsnm{Zhang}, \binits{F.}},
\bauthor{\bsnm{Xiao}, \binits{M.}},
\bauthor{\bsnm{Liu}, \binits{J.}},
\bauthor{\bsnm{Qu}, \binits{E.}}:
\batitle{Design of the guide tube calibration system for the {JUNO}
  experiment}.
\bjtitle{Journal of Instrumentation}
\bvolume{14}(\bissue{09}),
\bfpage{09005}--\blpage{09005}
(\byear{2019})
\doiurl{10.1088/1748-0221/14/09/t09005}
\end{barticle}
\endbibitem

\bibitem[\protect\citeauthoryear{Guo et~al.}{2021}]{Guo:2021ugw}
\begin{barticle}
\bauthor{\bsnm{Guo}, \binits{Y.}},
\bauthor{\bsnm{Zhu}, \binits{K.}},
\bauthor{\bsnm{Zhang}, \binits{Q.}},
\bauthor{\bsnm{Zhang}, \binits{F.}},
\bauthor{\bsnm{Meng}, \binits{Y.}},
\bauthor{\bsnm{Liu}, \binits{J.}},
\bauthor{\bsnm{Qu}, \binits{E.}}:
\batitle{Construction and simulation bias study of the guide tube calibration
  system for {JUNO}}.
\bjtitle{Journal of Instrumentation}
\bvolume{16}(\bissue{07}),
\bfpage{07005}
(\byear{2021})
\doiurl{10.1088/1748-0221/16/07/t07005}
\end{barticle}
\endbibitem

\bibitem[\protect\citeauthoryear{Zhu et~al.}{2019}]{USS_paper}
\begin{barticle}
\bauthor{\bsnm{Zhu}, \binits{G.-L.}},
\bauthor{\bsnm{Liu}, \binits{J.-L.}},
\bauthor{\bsnm{Wang}, \binits{Q.}},
\bauthor{\bsnm{Xiao}, \binits{M.-J.}},
\bauthor{\bsnm{Zhang}, \binits{T.}}:
\batitle{{Ultrasonic positioning system for the calibration of central
  detector}}.
\bjtitle{Nucl. Sci. Tech.}
\bvolume{30}(\bissue{1}),
\bfpage{5}
(\byear{2019})
\doiurl{10.1007/s41365-018-0530-x}
\end{barticle}
\endbibitem

\bibitem[\protect\citeauthoryear{Feldman and Cousins}{1998}]{Feldman_1998}
\begin{barticle}
\bauthor{\bsnm{Feldman}, \binits{G.J.}},
\bauthor{\bsnm{Cousins}, \binits{R.D.}}:
\batitle{Unified approach to the classical statistical analysis of small
  signals}.
\bjtitle{Physical Review D}
\bvolume{57}(\bissue{7}),
\bfpage{3873}--\blpage{3889}
(\byear{1998})
\doiurl{10.1103/physrevd.57.3873}
\end{barticle}
\endbibitem

\bibitem[\protect\citeauthoryear{Teng et~al.}{2022}]{Teng:2022usb}
\begin{barticle}
\bauthor{\bsnm{Teng}, \binits{D.}},
\bauthor{\bsnm{Liu}, \binits{J.-L.}},
\bauthor{\bsnm{Zhu}, \binits{G.-L.}},
\bauthor{\bsnm{Meng}, \binits{Y.}},
\bauthor{\bsnm{Zhang}, \binits{Y.-y.}},
\bauthor{\bsnm{Zhang}, \binits{T.}},
\bauthor{\bsnm{Luo}, \binits{K.}},
\bauthor{\bsnm{Li}, \binits{R.}},
\bauthor{\bsnm{Hui}, \binits{J.-Q.}}:
\batitle{{Low-radioactivity ultrasonic hydrophone used in positioning system
  for Jiangmen Underground Neutrino Observatory}}.
\bjtitle{Nucl. Sci. Tech.}
\bvolume{33}(\bissue{6}),
\bfpage{76}
(\byear{2022})
\doiurl{10.1007/s41365-022-01059-1}
\end{barticle}
\endbibitem

\bibitem[\protect\citeauthoryear{Berger et~al.}{2010}]{NIST_XCOM}
\begin{botherref}
\oauthor{\bsnm{Berger}, \binits{M.J.}},
\oauthor{\bsnm{Hubbell}, \binits{J.H.}},
\oauthor{\bsnm{Seltzer}, \binits{S.M.}}, et al.:
{XCOM: Photon Cross Sections Database}.
NIST Standard Reference Database 8 (XGAM),
  \url{https://physics.nist.gov/PhysRefData/Xcom/html/xcom1.html}
(2010)
\end{botherref}
\endbibitem

\bibitem[\protect\citeauthoryear{Agostinelli et~al.}{2003}]{geant4_paper}
\begin{barticle}
\bauthor{\bsnm{Agostinelli}, \binits{S.}}, \betal:
\batitle{{{GEANT4}--a simulation toolkit}}.
\bjtitle{Nucl. Instrum. Meth. A}
\bvolume{506},
\bfpage{250}--\blpage{303}
(\byear{2003})
\doiurl{10.1016/S0168-9002(03)01368-8}
\end{barticle}
\endbibitem

\bibitem[\protect\citeauthoryear{Zou et~al.}{2015}]{SNIPER}
\begin{barticle}
\bauthor{\bsnm{Zou}, \binits{J.H.}},
\bauthor{\bsnm{Huang}, \binits{X.T.}},
\bauthor{\bsnm{Li}, \binits{W.D.}},
\bauthor{\bsnm{Lin}, \binits{T.}},
\bauthor{\bsnm{Li}, \binits{T.}},
\bauthor{\bsnm{Zhang}, \binits{K.}},
\bauthor{\bsnm{Deng}, \binits{Z.Y.}},
\bauthor{\bsnm{Cao}, \binits{G.F.}}:
\batitle{{{SNiPER}: an offline software framework for non-collider physics
  experiments}}.
\bjtitle{J. Phys. Conf. Ser.}
\bvolume{664}(\bissue{7}),
\bfpage{072053}
(\byear{2015})
\doiurl{10.1088/1742-6596/664/7/072053}
\end{barticle}
\endbibitem

\bibitem[\protect\citeauthoryear{Collaboration}{2025}]{CPC_energy_resolution}
\begin{barticle}
\bauthor{\bsnm{Collaboration}, \binits{T.J.}}:
\batitle{Prediction of energy resolution in the {JUNO} experiment}.
\bjtitle{Chinese Physics C}
\bvolume{49}(\bissue{1}),
\bfpage{013003}
(\byear{2025})
\doiurl{10.1088/1674-1137/ad83aa}
\end{barticle}
\endbibitem

\bibitem[\protect\citeauthoryear{Abusleme et~al.}{2025}]{2025result1juno}
\begin{botherref}
\oauthor{\bsnm{Abusleme}, \binits{A.}},
\oauthor{\bsnm{Adam}, \binits{T.}}, et al.:
Initial performance results of the JUNO detector
(2025).
\url{https://arxiv.org/abs/2511.14590}
\end{botherref}
\endbibitem

\bibitem[\protect\citeauthoryear{and Abusleme et~al.}{2021}]{JUNO-bkg-paper}
\begin{botherref}
\oauthor{\bsnm{Abusleme}, \binits{A.}},
\oauthor{\bsnm{Adam}, \binits{T.}}, et al.:
Radioactivity control strategy for the {JUNO} detector.
Journal of High Energy Physics
\textbf{2021}(11)
(2021)
\doiurl{10.1007/jhep11(2021)102}
\end{botherref}
\endbibitem

\bibitem[\protect\citeauthoryear{Takenaka et~al.}{2025}]{Takenaka_2025}
\begin{botherref}
\oauthor{\bsnm{Takenaka}, \binits{A.}},
\oauthor{\bsnm{Chen}, \binits{Z.}},
\oauthor{\bsnm{Freegard}, \binits{A.}},
\oauthor{\bsnm{Huang}, \binits{J.}},
\oauthor{\bsnm{Hui}, \binits{J.}},
\oauthor{\bsnm{Lai}, \binits{H.}},
\oauthor{\bsnm{Li}, \binits{R.}},
\oauthor{\bsnm{Liao}, \binits{Y.}},
\oauthor{\bsnm{Liu}, \binits{J.}},
\oauthor{\bsnm{Meng}, \binits{Y.}},
\oauthor{\bsnm{Morton-Blake}, \binits{I.}},
\oauthor{\bsnm{Xiang}, \binits{Z.}},
\oauthor{\bsnm{Zhang}, \binits{P.}}:
Neutron source-based event reconstruction algorithm in large liquid
  scintillator detectors.
The European Physical Journal C
\textbf{85}(10)
(2025)
\doiurl{10.1140/epjc/s10052-025-14808-4}
\end{botherref}
\endbibitem

\bibitem[\protect\citeauthoryear{Abusleme et~al.}{2026}]{2026result1juno}
\begin{barticle}
\bauthor{\bsnm{Abusleme}, \binits{A.}},
\bauthor{\bsnm{Adam}, \binits{T.}},
\bauthor{\bsnm{{The JUNO Collaboration}}}:
\batitle{Measurement of reactor neutrino oscillation with the first {JUNO}
  data}.
\bjtitle{Nature}
\bvolume{654}(\bissue{8118}),
\bfpage{343}--\blpage{348}
(\byear{2026})
\end{barticle}
\endbibitem

\end{thebibliography}
	
\clearpage
\appendix	
\renewcommand{\thetable}{A\arabic{table}}   
\setcounter{table}{0}                       
\begin{sidewaystable}[!ht]
	\centering
	\caption{Measured radioactivities of materials and parts in the calibration systems. All upper limits are quoted at 90\% C.L. For the NAA results reported as nuclide mass fractions (g/g), 1~g/g corresponds to $1.24\times10^{7}$~Bq/kg for $^{238}$U, $4.06\times10^{6}$~Bq/kg for $^{232}$Th, and $2.65\times10^{8}$~Bq/kg for $^{40}$K.}
	\renewcommand{\arraystretch}{1.4}
	\tiny
	\setlength{\tabcolsep}{2pt}
	\begin{tabular}{ccccccccccc}
		\hline
		\multirow{2}{*}{System} & \multirow{2}{*}{Detector} & \multirow{2}{*}{Material} 
		& \multirow{2}{*}{Supplier}
		& \multirow{2}{*}{Quantity} & \multirow{2}{*}{Unit} & \multicolumn{4}{c}{Activity}  \\
		& & & & &  & {$^{238}\mathrm{U}$} & {$^{232}\mathrm{Th}$} & 
		{$^{40}\mathrm{K}$} & {$^{60}\mathrm{Co}$} & {$^{137}\mathrm{Cs}$}\\ 
		\hline
		
		USS cable
		& \begin{tabular}{@{}c@{}}CJPL\\SJTU HPGe\end{tabular}
		& PTFE and copper
		& Pasterneck
		& 1.45~kg
		& mBq/kg
		& $3.42 \pm 0.47$
		& $1.94 \pm 0.49$
		& $(1.74 \pm 0.38)\times 10^{1}$
		& $< 3.70\times 10^{-1}$ 
		& $< 1.30\times 10^{-1}$\\
		\hline
		
		USS receiver
		& \begin{tabular}{@{}c@{}}CJPL\\SJTU HPGe\end{tabular}
		& Ni, Epoxy, PCB
		& Customized
		& 10~piece
		& mBq/piece
		& $< 3.30$
		& $< 2.80$
		& $< 1.90\times 10^{1}$
		& $< 8.60\times 10^{-1}$
		& $< 7.90\times 10^{-1}$ \\
		\hline
		
		CLS cable
		& \begin{tabular}{@{}c@{}}CJPL\\SJTU HPGe\end{tabular}
		& PTFE and SS
		& Fengshuo
		& 0.27~kg
		& mBq/kg		
		& $< 6.47\times 10^{1}$
		& $< 7.92\times 10^{1}$
		& $< 5.10\times 10^{2}$
		& $< 3.75\times 10^{1}$
		& $< 4.08\times 10^{1}$ \\
		\hline
		
		CLS Anchor
		& \begin{tabular}{@{}c@{}}Milano\\NAA\end{tabular}
		& PTFE
		& Sanxin Inc.
		& 6.4~kg
		& g/g
	
		& $< 1.20\times10^{-12}$
		& $(1.00 \pm 0.20)\times10^{-11}$
		& $(1.50 \pm 0.05)\times10^{-12}$
	    & --
	    & -- \\
		\hline
		
		Wear-resisting block
		& \begin{tabular}{@{}c@{}}Milano\\NAA\end{tabular}
		& PTFE
		& Sanxin Inc.
		& 6.9~kg
		& g/g
		& $< 1.20\times10^{-12}$
		& $(1.00 \pm 0.20)\times 10^{-11}$
		& $(1.50 \pm 0.05)\times 10^{-12}$ 
	    & --		
		& --\\
		\hline
		
		GTCS Tube
		& \begin{tabular}{@{}c@{}}Milano\\NAA\end{tabular}
		& PTFE
		& Jianwei
		& 20~kg
		& g/g
		& $< 1.65\times 10^{-12}$
		& $< 3.00\times 10^{-12}$
		& $(4.70 \pm 0.10)\times 10^{-10}$ 
	    & --		
		& --\\
		\hline
		
		GTCS cable
		& \begin{tabular}{@{}c@{}}CJPL\\SJTU HPGe\end{tabular}
		& PTFE and SS
		& Fengshuo
		& 0.33~kg
		& mBq/kg
		& $< 3.90\times 10^{1}$
		& $< 4.32\times 10^{1}$
		& $< 1.40\times 10^{2}$
		& $(2.85 \pm 0.11)\times 10^{2}$ 
	    & --\\
		\hline
		
		GTCS sensor
		& \begin{tabular}{@{}c@{}}CJPL\\SJTU HPGe\end{tabular}
		& SS, PVC, and PBT
		& Omron
		& 10~pieces
		& mBq/piece
		& $(1.11 \pm 0.03)\times 10^{2}$
		& $(1.84 \pm 0.05)\times 10^{2}$
		& $(3.83 \pm 0.21)\times 10^{2}$
		& $< 2.85$
		& $< 4.45$\\
		\hline		
	\end{tabular}
	\label{tab:Radio_measure}
\end{sidewaystable}

\end{document}